\documentstyle[prl,aps,multicol,subfigure,psfig,subeqnar]{revtex}

\begin{document}

\title{Dark soliton creation in Bose-Einstein condensates}
\author{L.~D. Carr$^{1}$\cite{byline},
J. Brand$^{2}$, S. Burger$^{3}$\cite{address}, and A. Sanpera$^{4}$\\}
\address{$^{1}$Department of Physics, University of Washington, 
         Seattle, WA 98195-1560, USA\\}
\address{$^{2}$Department of Chemistry, University of Washington, 
         Seattle, WA 98195-1700, USA\\}
\address{$^{3}$Institut f\"ur Quantenoptik, Universit\"at Hannover, 30167 Hannover, Germany\\}
\address{$^{4}$Institut f\"ur Theoretische Physik, Universit\"at Hannover, 30167 Hannover, Germany\\}
\maketitle

\date{\today}

\begin{abstract}
It is demonstrated that stable, standing dark solitons can be created in current dilute--gas Bose-Einstein condensate experiments by the proper combination of phase and density engineering.  Other combinations result in a widely controllable range of grey solitons.  The phonon contribution is small and is calculated precisely.  The ensuing dynamics should be observable {\it in situ}, i.e. without ballistic expansion of the condensate. 
\end{abstract}

\pacs{}

\begin{multicols}{2}

The observation of solitons in Bose-Einstein condensates (BEC's) brings together two heretofore disparate fields of study: quantum degenerate gases~\cite{dalfovo1} and exactly integrable nonlinear systems~\cite{zakharov2}.  Recently three-dimensional, solitonic structures have been observed in weakly--interacting atomic gas BEC's for the first time~\cite{burger1,denschlag1}.  As fundamentally nonlinear, collective excitations of a macroscopic quantum wave they are intriguing in their own right.  However, standing solitons in one dimension, which are expected to have a wealth of special properties~\cite{kivshar3}, have not yet been created.

The Gross-Pitaevskii equation which describes the mean field dynamics of a dilute BEC at low temperatures~\cite{dalfovo1} reduces to a one-dimensional nonlinear Schr\"odinger equation (NLS) when the transverse dimensions of the condensate are on the order of its healing length and its longitudinal dimension is much longer than its transverse ones~\cite{carr15}.  This is termed the \emph{quasi-one-dimensional} (quasi-1D) regime of the Gross-Pitaevskii equation.  If the transverse dimensions of the condensate are much less than the healing length then the Gross-Pitaevskii equation no longer applies and other physics ensue~\cite{petrov2}.  The recent confinement of a BEC in a hollow blue--detuned laser beam demonstrates that the quasi-1D regime is experimentally realizable~\cite{bongs1}.  In three dimensions solitonic structures are unstable to transverse modulations but in quasi-1D they are stable, and are in fact the stationary states of the NLS~\cite{carr22}.  

The extent to which quantum fluctuations affect the BEC is an outstanding theoretical question~\cite{morgan3}.  As dark solitons are solutions to the mean field theory, substantial instability of dark solitons at low temperature in quasi-1D is an experimental measure of such higher order effects.  The present experimental technique for creating solitonic structures is based on phase engineering alone, makes additional transient density waves, and cannot be used to make a single standing dark soliton.  Observations have been made in three-dimensional harmonic geometries or at temperatures for which about 10\% of the atoms remained uncondensed.  Time scales for observation are subsequently on the order of 10 ms, whereas the lifetime of the BEC is on the order of seconds, precluding the likelihood of observing the effect of quantum fluctuations.  Finally, the length scale of these solitonic structures, i.e. the healing length, is so small that {\it in situ} observation of their dynamics is not possible and the condensate must first be expanded by turning off the confining potential.  In the following, a method which remedies all of these difficulties is presented.

The NLS may be written~\cite{carr22} in the form
\begin{equation}
\label{eqn:nls}
[\,-\xi^2\partial_{xx}+ L|\psi(x,t)|^2+ V(x)\,]\,\psi(x,t) = i \partial_t\psi(x,t) \, ,
\end{equation}
where $\psi$ is the macroscopic quantum wavefunction, $L$ is a longitudinal confinement length, $V(x)=(2m\xi^2/\hbar^2)V_o(x)$ is a rescaling of a confining potential $V_o(x)$, $\xi\equiv(8\pi\bar{n}a)^{-1/2}$ is the healing length, $a$ is the s-wave scattering length for binary atomic interactions, and $\bar{n}$ is the mean number density.  $\psi$ is normalized to 1 and has units of $L^{-1/2}$, $x$ has units of length, and $t\equiv(2m\xi^2/\hbar)t_o$ is dimensionless, where $t_o$ has units of time.  

In particular, we consider the case of $^{87}$Rb confined in a box-like potential, with $L=100\,\mu$m and a transverse length of $L_t=10\,\mu$m.  For $N=1.1\times 10^4$ atoms and $\bar{n}\equiv N/(LL_t^2)$ the effective healing length~\cite{footnote1} is 2.5 $\mu$m.  As the length scale of solitons is $\sim 2\,\xi$ and the wavelength of the imaging radiation is $\sim 0.5\,\mu$m, this experimentally realizable configuration ensures observability without needing to ballistically expand the condensate.  It also ensures the possibility of phase and density engineering of structures on the scale of the healing length by means of the dipole potential of laser fields.  A key point in this choice of parameters is the use of a box-like, rather than harmonic, confining potential~\cite{footnote2}.  For harmonic confinement the Thomas-Fermi radius~\cite{dalfovo1} of the condensate scales as $N^{1/5}$, $\bar{n}\propto N/N^{3/5}$, and therefore $\xi\propto N^{-1/5}$, whereas for box-like confinement $\xi\propto N^{-1/2}$.

Neglecting the effect of $V(x)$ and writing the wavefunction as $\psi(x,t) = \sqrt{\rho(x,t)}\exp[i\phi(x,t)-i\mu t]$, the single
%
%
\begin{figure}
\centerline{\psfig{figure=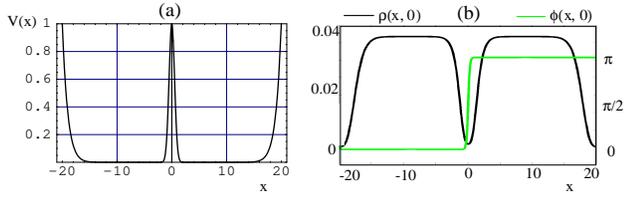,width=83mm}}
\begin{center}
\begin{minipage}{83mm}
\caption{(a) A combination of a box-like potential and a tightly focused, blue--detuned laser beam is used to engineer the density.  (b) The resulting wavefunction is phase engineered with a second, far--detuned laser beam, resulting in an initial state very close to a standing dark soliton.
}
\label{fig:1}
\end{minipage}
\end{center}
\end{figure}
%
%
\noindent soliton solution to Eq.~(\ref{eqn:nls}) takes the form
\begin{subeqnarray}
\label{eqn:soliton}
\rho(x,t) &=& \frac{1}{L}\left\{1-2\gamma^2\xi^2\,sech^2[\gamma\,(x-c\,t)]\right\} \\
\phi(x,t) &=& \tan^{-1}\left\{\frac{2\xi^2\gamma}{c}\,tanh[\gamma\,(x-c\,t)]\right\} +\frac{\pi}{2} \\
{\rm where}&&2\gamma^2\xi^2+c^2/(2\xi^2)=1\, .
\end{subeqnarray}
For $L\gg\xi$ Eqs.~(\ref{eqn:soliton}) are an excellent approximation away from the walls.  The single particle density $\rho(x,t)$ is constant except over a density notch of width $1/\gamma$, and the phase changes sharply and monotonically across the notch.  The chemical potential $\mu = 1$ in these units and $c$ is the speed of the soliton.  The constraint (\ref{eqn:soliton}c) links the soliton velocity $c$ to its depth $2\xi^2\gamma^2$ and thus leaves only a single free parameter for the soliton solution.  In the case $c=0$, $1/\gamma=\sqrt{2}\xi$ and the soliton forms a node, is a stationary solution to Eq.~(\ref{eqn:nls}), and is called a \emph{dark soliton}.  In the case $0<c<c_{s}$, where $c_{s}\equiv\sqrt{2}\xi$ is the Bogoliubov sound speed, the soliton is a moving density notch and is called a \emph{grey soliton}.  In the case $c\rightarrow c_{s}^{-}$, $1/\gamma$ diverges and the soliton depth approaches zero.  Upon reinsertion of constants it is found that $c_{s}=\sqrt{4\pi\bar{n}a}\,\hbar/m$.  The total phase difference across the notch~\cite{reinhardt1} is $\Delta\phi=2\tan^{-1}[\sqrt{1-(c/c_s)^2}/(c/c_s)]$.

As is apparent from Eq.~(\ref{eqn:soliton}), a soliton requires both a nontrivial phase and a nontrivial density profile.  The technique of phase engineering used to make solitonic structures in previous experiments~\cite{burger1,denschlag1} imprinted a constant phase on one half of a nearly uniform condensate.  This resulted in a combination of transient density waves and one or more grey solitons.  We propose instead the following scheme.  

Atoms are condensed into a box-like trap with a sharply focused, blue--detuned laser beam in the center of the long trap axis.  Due to a large laser detuning from atomic resonances, spontaneous processes can be neglected on the relevant time scales.  Thus the effect of the laser beam on the condensate atoms can be described by the optical dipole potential, which is proportional to the local light intensity.  In Fig.~\ref{fig:1}(a) is shown the potential used as a model, a gaussian with a width on the order of the healing length, i.e. $2\Delta x_v\sim 2.5\mu$m, steep walls at
%
%
\begin{figure}
\centerline{\psfig{figure=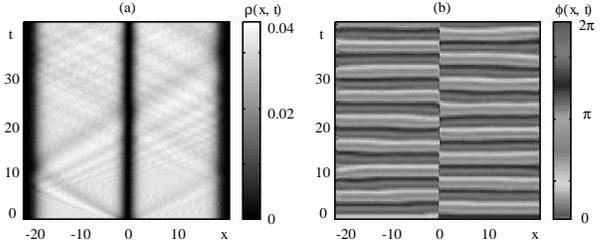,width=83mm}}
\begin{center}
\begin{minipage}{83mm}
\caption{A standing dark soliton is created with the proper combination of phase and density engineering.  In addition to the central dark soliton, characterized by a $\pi$ phase difference across a density notch which forms a node, a spray of phonons is emitted at the speed of sound ($c_s = \sqrt{2}$) to either side.  A second, shallow soliton, emitted to the left at nearly $c_s$, offsets the effect of the finite slope of the initial phase due to the diffraction--limited fall-off of the phase--engineering laser beam.
}
\label{fig:2}
\end{minipage}
\end{center}
\end{figure}
%
%
\noindent $\pm L/2\,$, and a height on the order of the chemical potential $\mu=1\,$:
\begin{equation}
V(x)= \exp\left(\frac{-x^2}{2\,(\Delta x_v)^2}\right)+\left(\frac{2\,x}{L}\right)^{20} \, .
\label{eqn:potential}
\end{equation}
In Fig.~\ref{fig:1} and in all simulations the parameters $\xi=1$ and $L=40$ were used.  Note that it should also be possible to create a density notch by adiabatically ramping the intensity of a laser into an initially uniform BEC.  The resulting deformation dips to $\sim$1\% of the maximum density, or $\sim$10\% of the maximum amplitude, as may be seen in Fig.~\ref{fig:1}(b).  Because the scale of variation of $V(x)$ is $\lesssim\xi$ the kinetic energy term in Eq.~(\ref{eqn:nls}) cannot be neglected and the response of the condensate is not in the Thomas-Fermi regime.  It is therefore obtained numerically by imaginary time relaxation of Eq.~(\ref{eqn:nls}) with Eq.~(\ref{eqn:potential}).

At $t=0$ the focused laser beam is switched off and a second, far--detuned laser pulse of uniform intensity distribution is shined on one half of the condensate.  The pulse duration is chosen to be shorter than the correlation time of the condensate, $t_c=\xi/c_s$.  This ensures that the light field changes only the phase distribution and not the density distribution of the BEC.  In the simulation it therefore suffices to switch on a controllable phase $\phi(x)$ instantaneously~\cite{burger1}.
\begin{equation}
\label{eqn:phase}
\phi(x)=\Delta\phi\,\tanh(2x/\Delta x_{\phi})
\end{equation}
was used as a model, which represents the diffraction--limited fall-off of the laser beam over the notch, with a width of $\Delta x_{\phi}$ and a total phase difference of $\Delta\phi$, as may be seen in Fig.~\ref{fig:1}(b), where the parameters $\Delta\phi=\pi$ and $\Delta x_{\phi}=\xi\sim 2.5\,\mu$m have been used.

The resulting notch is both density and phase engineered.  In Fig.~\ref{fig:2} is shown its evolution.  For $^{87}$Rb there 
%
%
\begin{figure}
\centerline{\psfig{figure=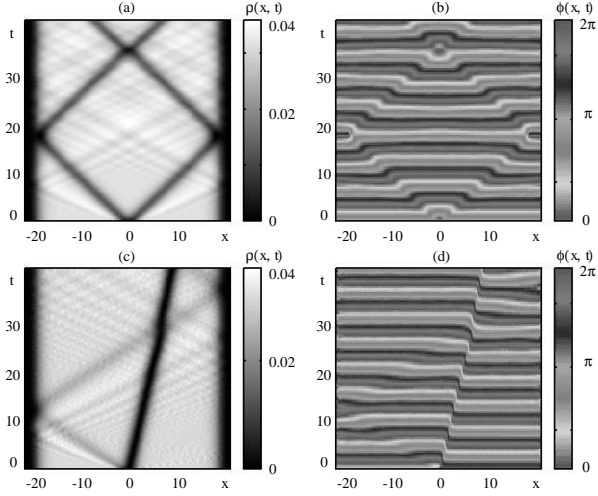,width=83mm}}
\begin{center}
\begin{minipage}{83mm}
\caption{A variety of physically intriguing possibilities arise from variations in the technnique used to create a standing dark soliton.  (a)-(b) The use of density engineering alone creates a pair of equal and opposite grey solitons with a widely controllable range of velocites.  (c)-(d) Phase engineering with a phase different from $\pi$ creates an asymmetric grey soliton pair.  Note the spatial shift inherent in soliton--soliton interactions.
}
\label{fig:3}
\end{minipage}
\end{center}
\end{figure}
%
%
\noindent are 2.27 ms per simulation time unit~\cite{carr22}.  When $\Delta\phi=\pi$ was used it was found that the soliton drifted slowly to the left, i.e. was not completely dark.  A true standing dark soliton requires a step function in the phase.  Therefore to counter the effect of the diffraction--limited fall-off of the second laser beam, $\Delta\phi=1.05\times\pi$ was used.  This creates a second, shallow grey soliton which carries away the small momentum of the first without appreciably deforming the density.  Thus a standing dark soliton is successfully created.

As is apparent in Fig.~\ref{fig:2}, some portion of the initial notch is radiated as phonons.  The apportionment of the total energy into phonons and solitons may be calculated precisely.  Because the NLS is nonlinear it is not possible to build any particular solution by a summation over phonon modes.  As phonons and solitons are the fundamental solution types of the 1D NLS it is therefore useful to single out their contribution to the dynamics.  The NLS has a denumerably infinite number of integrals of motion~\cite{kivshar3}.  In the context of the BEC the first three correspond to normalization, energy, and momentum.  For the simulations studied herein it suffices to consider energy.

The second integral of motion of Eq.~(\ref{eqn:nls}) is
\begin{equation}
\label{eqn:integralofmotion}
I_2=\int_{-L/2}^{L/2} dx \left[\xi^2|\partial_x \psi|^2 +\frac{L}{2}|\psi|^4 + V(x)|\psi|^2\right ] \, .
\end{equation}
Assuming $L\gg\xi$, the energy of a single soliton is 
\begin{equation}
\label{eqn:solitonintegralofmotion}
E_s\simeq N\int_{-\infty}^{\infty} dx \left [\xi^2|\partial_x \psi|^2 +\frac{L}{2}\left (|\psi|^2-\frac{1}{L}\right )^2\right ]\, ,
\end{equation}
where Eq.~(\ref{eqn:integralofmotion}) has been renormalized to subtract out the constant background and the potential $V(x)$ has been assumed to be constant in the region of the soliton.  The factor of $N$ follows from the use of a single, rather than $N$, particle density for $|\psi(x,t)|^2$.  Equations~(\ref{eqn:soliton}) may then be used to obtain the energy of a single soliton:

\begin{equation}
\label{eqn:solitonenergy}
E_s \simeq \frac{N}{L}\,\frac{4}{3}\, c_s\,|\sin(\Delta\phi/2)|^3 \, .
\end{equation}
$E_s$ is dimensionless and depends on the line density $N/L$.  The rescaling $[\hbar^2/(2m\xi^2)]E_s\propto (N/L)^{3/2}$ reinstates the units of energy.  For our choice of parameters the difference between the approximation and an exact expression for $E_s$ is negligible and not observable in simulations.

Conservation of the total energy of the system may be expressed as $E_d+ E_{o} = E_s + E_p + E_{o}$, where $E_d$, $E_{o}$, and $E_{p}$ are the energies of the initial deformation, the undeformed background ground state, and the phonons, respectively.  By use of Eq.~(\ref{eqn:integralofmotion}) $E_d$ and $E_o$ may be determined in the simulation.  $E_s$ may be calculated from Eq.~(\ref{eqn:solitonenergy}). With this method it was determined that the total phonon contribution in Fig.~\ref{fig:2} is 39\% of the deformation energy $E_d$ and the second soliton carries away only 1.5\% of $E_d$.  However, the phonons spread out over the whole box while the 60\% of the energy carried by the dark soliton is localized in a region of width $\sim 2\xi=2$.  Therefore this method of dark soliton creation is highly efficient and phonons are essentially small fluctuations in the density.

It is important to explore the robustness of the proposed technique.  To what extent does it depend on the parameters of Eqs.~(\ref{eqn:potential})-(\ref{eqn:phase})?  To this end many simulations were performed which varied $\Delta x_v$, $\Delta x_{\phi}$, and $\Delta \phi$.  It was found that $2\Delta x_v\leq\xi$ was optimal.  Note that the full width of the gaussian is $2\Delta x_v$.  $2\Delta x_v=2\xi$ produced an additional pair of very shallow grey solitons which would not be observable in experiments.  $\Delta x_{\phi}\leq\xi$ was sufficient to produce a standing dark soliton while $L\gg\Delta x_{\phi}>\xi$ set the soliton in motion. Finally $\Delta \phi\sim\pi$ was required, as expected.  It was also found that the amount of the deformation energy transformed into phonons was consistently about 40\% to 50\%.  Outside of these parameter ranges a number of additional intriguing possibilities arose.  In the following a few such possibilities are illustrated.  

If density engineering alone is used, a pair of equal and opposite grey solitons emerge.  This effect is illustrated in Fig.~\ref{fig:3}(a)-(b).  The velocities of the grey soliton pair may, by use of Eq.~(\ref{eqn:solitonenergy}), be used to determine the total energy of the phonons.  It was found in simulations that within the range of $2\Delta x = $ 0.5 to 2.5 from 40\% to 70\% of the deformation energy was converted into phonons and the remaining energy was converted into a symmetric grey soliton pair.  In general a lower energy notch resulted
%
%
\begin{figure}
\centerline{\psfig{figure=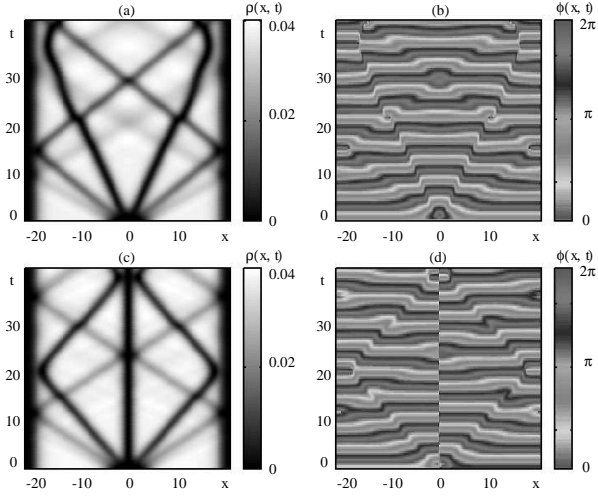,width=83mm}}
\begin{center}
\begin{minipage}{83mm}
\caption{(a)-(b) Density engineering a notch wider than the healing length creates a symmetric fan of grey solitons.  (c)-(d) If phase engineering is added a dark soliton is created in the center of the fan.
}
\label{fig:4}
\end{minipage}
\end{center}
\end{figure}
%
%
\noindent in slower solitons, with a lower limit on the soliton velocity of $\sim 0.45 c_s$.  In previous solitonic--structure experiments, which used phase engineering alone~\cite{dobrek1}, the velocities were 0.54$c_s$ to 0.76$c_s$ and 0.68$c_s$ to $c_s$, respectively~\cite{burger1,denschlag1}.  By the method of density engineering a wide range of velocities from $0.45 c_s$ to $c_s$ may be obtained by adjusting the width or depth and thereby the energy of the initial deformation.

Another variation on the technique, which uses both density and phase engineering and a phase difference of $\Delta\phi\neq\pi$, creates an asymmetric grey soliton pair.  60\% to 70\% of the deformation energy is radiated as phonons.  For the phase profile of Eq.~(\ref{eqn:phase}), $\pi < \Delta\phi < 2\pi$ results in a faster soliton to the right.  $0 < \Delta\phi < \pi$ results in a faster soliton to the left, as is to be expected from the constant phase offset inherent in the wavefunction.  The total phase difference across the two solitons adds up to $\Delta \phi$.  Such asymmetric interactions are useful in observing the delay inherent in soliton interactions~\cite{carr29}.

If again density engineering alone is used, but with too large a notch, i.e. $\Delta x_v > \xi$, multiple pairs of equal and opposite grey solitons emerge, as has also been predicted elsewhere~\cite{reinhardt1}.  If phase engineering is added to this scenario a central, dark soliton is created in the middle of the trap with equal and opposite pairs of grey solitons fanning outwards to either side.  These cases are shown in Fig.~\ref{fig:4}.

In general the consistent 40\%-70\% of the deformation energy radiated into phonons is representative of the difference in form of the density deformation which results from the gaussian in Eq.~(\ref{eqn:potential}) and the sech in Eq.~(\ref{eqn:soliton}a).  A different density notch could be expected to result in a different apportionment of energy.  It should also be noted that solitons are robustly stable structures~\cite{kivshar1} and, though not shown here, the introduction of 10\% stochastic noise into the simulations had no observable effect on the results.


In conclusion, a method of creating a single, standing dark soliton in a gaseous Bose-Einstein condensate has been presented.  This experimentally realizable method combines phase and density engineering and requires one or two lasers and a quasi-one-dimensional trap.  Variations result in the creation of grey solitons with a widely controllable range of velocities.

We thank Klaus Sengstock and William Reinhardt for helpful discussions.  This work was supported by National Science Foundation Grant CHE97-32919 and by SFB407 of the {\it Deutsche Forschungsgemeinschaft}.  J. Brand acknowledges support from the Alexander von Humboldt Foundation through a Feodor Lynen Research Fellowship.


\begin{thebibliography}{10}

\bibitem[*]{byline} to whom correspondence should be addressed

\bibitem[\dag]{address} Current address: INFM - L.E.N.S., Largo E. Fermi, 2, 50125 Firenze, Italia

\bibitem{dalfovo1}
F. Dalfovo, S. Giorgini, L.~P. Pitaevskii, and S. Stringari, Rev. Mod. Phys.
  {\bf 71},  463  (1999).

\bibitem{zakharov2}
V.~E. Zakharov and A.~B. Shabat, Sov. Phys. JETP {\bf 37},  823  (1973).

\bibitem{burger1}
S. Burger {\it et~al.}, Phys. Rev. Lett. {\bf 83},  5198  (1999).

\bibitem{denschlag1}
J. Denschlag {\it et al.}, Science {\bf 287},  97  (2000).

\bibitem{kivshar3}
Y.~S. Kivshar, Physics Reports {\bf 298},  81  (1998).

\bibitem{carr15}
L.~D. Carr, C.~W. Clark, and W.~P. Reinhardt, Phys. Rev. A {\bf 62} 063610-1 (2000).

\bibitem{petrov2}
D.~S. Petrov, G.~V. Shylapnikov, and J.~T.~M. Walraven, Phys. Rev. Lett. {\bf
  85},  3745  (2000).

\bibitem{bongs1}
K. Bongs {\it et~al.}, e-print cond-mat/0007381 (2000).

\bibitem{dobrek1}
L. Dobreck {\it et al.}, Phys. Rev. A {\bf 60}, 3381 (1999).

\bibitem{carr22}
L.~D. Carr, M.~A. Leung, and W.~P. Reinhardt, J. Phys. B {\bf 33},  3983
  (2000).

\bibitem{morgan3}
S.~A. Morgan, e-print cond-mat/9911278 (1999).

\bibitem{footnote1}
The effective healing length takes into account the transverse dimensions of the condensate by making the approximation of separation of variables and integrating over the transverse ground states, and is discussed in detail elsewhere~\cite{carr22}.

\bibitem{footnote2}
A hollow blue-detuned laser beam can be made to have a wide range of transverse confinement potentials by the use of higher modes.  In particular, the quasi-1D experiment of Ref.~\cite{bongs1} is capable of producing $V(\rho)=\rho^{20}$, where $\rho$ is the radius.

\bibitem{reinhardt1}
W.~P. Reinhardt and C.~W. Clark, J. Phys. B. {\bf 30},  L785  (1997).

\bibitem{carr29}
S. Burger, L.~D. Carr, P. Ohberg, and A. Sanpera, Phys. Rev. A, to be submitted (2000).

\bibitem{kivshar1}
Y.~S. Kivshar and X. Yang, Phys. Rev. E {\bf 49},  1657  (1994).

\end{thebibliography}

\end{multicols}
\end{document}